\documentclass[10pt,a4paper]{article} % 10pt is ignored!
\usepackage{graphicx}
\usepackage{amsfonts,amsmath,amssymb}
\usepackage{hyperref}
\usepackage{cite}
\newcommand{\bse}{\begin{subequations}}
\newcommand{\ese}{\end{subequations}}

\begin{document}
\title{\bf Frame dependence of heavy-quark observables in a rotating strongly coupled plasma}

\date{}
\maketitle
\vspace*{-0.3cm}
\begin{center}
{\bf Leila Shahkarami$^{1}$, Farid Charmchi$^{2}$}\\
\vspace*{0.3cm}
{\it {School of Physics, Damghan University, Damghan 36716-45667, Iran
}} \\
\vspace*{0.3cm}
{\it  {${}^1$l.shahkarami@du.ac.ir}, {${}^2$farid.charmchi@gmail.com} }
\end{center}

\begin{abstract}
In genuinely rotating holographic backgrounds, Wilson-loop observables are usually constructed for a quark--antiquark pair static with respect to the rotating plasma. We investigate how this choice affects the heavy-quark sector by comparing two descriptions of the same rotating geometry: a corotating frame, in which the pair is static with respect to the plasma, and a static frame, in which it is static with respect to the static observer. We consider an equal-rotation Myers--Perry black hole in asymptotically global AdS and study Wilson loops oriented longitudinally and transversely to the rotation. We analyze the effective string tension, physical quark--antiquark separation, and heavy-quark potential at fixed thermodynamic temperature measured in the corresponding frame. We find a strong frame dependence of the rotational response. In the corotating frame, the longitudinal and transverse configurations respond qualitatively differently: rotation can either decrease or increase the separation and effective string tension, depending on orientation and, for the latter, on the bulk position and parameter range. The heavy-quark potential increases with angular velocity in both orientations. In the static frame, by contrast, the effective tension, physical separation, and maximal separation decrease with angular velocity in both orientations, while the heavy-quark potential increases. Thus, the anisotropic response is substantially stronger in the corotating frame, both qualitatively and quantitatively. In both frames, rotational modifications generally become less pronounced at higher temperature. These results show that the frame in which the quark pair is static is an essential part of the physical specification of Wilson-loop observables in rotating holographic systems.
\end{abstract}
Keywords: Rotating quark-gluon plasma; Wilson loop; Heavy quark potential.
%$$$$$$$$$$$$$$$$$$$$$$$$$$$$$$$$$$$$$$$$$$$$$$$$$$$$$$$$$$$$$$$$$$$$$$$$$
\section{Introduction}

Noncentral relativistic heavy-ion collisions generate a large orbital angular momentum \cite{angularm} and can produce strongly interacting matter with substantial vorticity. The observation of global polarization of $\Lambda$ and $\bar{\Lambda}$ hyperons by the STAR Collaboration \cite{star2017}, together with subsequent theoretical and experimental studies of polarization and spin--orbital-angular-momentum coupling \cite{Becattini2020,alice,star2023}, has established vorticity as an important physical property of the quark--gluon plasma (QGP) created in such collisions. The large angular momentum carried by the plasma makes it important to understand how rotation modifies the properties of strongly coupled matter. These developments have motivated extensive theoretical studies of rotating QCD matter using effective approaches, lattice simulations, and holographic methods. Among the many important questions in this field is how rotation modifies the thermodynamics and phase structure of strongly interacting matter and, in particular, how it affects the interaction and dissociation of heavy quark--antiquark pairs.

A lattice formulation of QCD in a rotating frame was introduced in \cite{latticerotating1st}, providing a framework in which rotation is incorporated through a non-inertial gravitational background and, as a first application, allowing the angular momentum carried by gluons and quarks to be studied. Subsequent lattice simulations of rotating gluodynamics found that the confinement--deconfinement transition temperature increases with angular velocity \cite{lattice2021}. Lattice simulations with dynamical quarks further showed that the gluonic and fermionic sectors respond to rotation with opposite shifts of the critical temperature, while their combined effect in QCD leads to an increasing pseudocritical temperature \cite{lattice2023}. An independent lattice study of rotating QCD with dynamical quarks likewise found that real rotation enhances confinement and chiral symmetry breaking, corresponding to an increase of the associated transition temperatures \cite{lattice3}. These results contrast with a substantial body of effective-model and holographic studies, which generally find that real rotation lowers the chiral or confinement--deconfinement transition temperature \cite{effective1,effective2,effective3,effective4,effective5,effective6,effective7,effective8,effective9,huang,transitionholo1,transitionholo2,boostholo,transitionholo3,transitionholo4,transitionholo5}. More recent approaches, however, have shown that this trend can be reversed when the response of the gluonic sector is incorporated: effective-model calculations constrained by lattice results and holographic constructions with a fully backreacted anisotropic background both find an increase of the transition temperatures with real rotation \cite{correctphase1,correctphase2}. Thus, the rotational modification of the phase structure is sensitive to how rotation is implemented and to the response of the different degrees of freedom in the strongly interacting system.

A particularly important development in this context is the recent observation that the apparent disagreement between holographic and lattice-QCD results for the confinement--deconfinement transition can be related to the choice of observer. In \cite{Braga2026}, the rotating plasma was described by an equal-rotation Myers--Perry black hole \cite{MPBH}, and the transition temperature was evaluated both for a static observer and for an observer corotating with the plasma. The latter choice was motivated by the fact that lattice-QCD calculations are performed in a reference frame comoving with the plasma \cite{latticerotating1st,lattice2021,lattice2023,lattice3,lattice4}. The two descriptions give opposite rotational trends for the critical temperature: an increasing trend is found for the temperature measured by a corotating observer, in contrast to the decreasing behavior obtained in the static frame. This difference is associated with the local temperature measured by the observer and its relation to the global thermodynamic temperature through the Tolman--Ehrenfest relation \cite{Tolman1930}, which was applied for the first time in holography in \cite{Braga2026} to distinguish thermodynamic measurements associated with different observers. The analysis showed that the choice of observer can qualitatively affect the rotational dependence of the transition temperature. It therefore raises a natural question for nonlocal observables: does the frame in which the quark--antiquark pair is held fixed likewise affect the rotational response of the Wilson loop and the heavy-quark observables extracted from it?

The heavy-quark potential is naturally defined through a temporal Wilson loop, and holography, through the gauge/gravity correspondence \cite{adscft1,adscft2,adscft3,adscft4}, provides a direct description in terms of the classical Nambu--Goto action of a string whose endpoints represent the heavy quark and antiquark \cite{Maldacenawilson}. Several holographic studies have investigated heavy-quark potentials and related observables in rotating backgrounds. These include genuinely rotating asymptotically AdS black holes, such as Kerr--AdS and Myers--Perry--AdS geometries \cite{KerrAdSwilson1,MPwilson1}, as well as rotating backgrounds constructed from nonrotating geometries through coordinate or Lorentz transformations \cite{boost1,boost2,boost3,boost4,global}. Within these constructions, the effects of rotation on heavy-quark binding, screening, string breaking, and related quantities have been studied, with some works also examining the dependence on the orientation of the quark--antiquark pair relative to the rotation axis \cite{MPwilson1,boost1,boost3,global}. In most of these studies, however, the frame with respect to which the quark and antiquark are held fixed is not treated as an independent physical specification; in particular, the question of how Wilson-loop observables change when the pair is held fixed in a frame corotating with the rotating plasma has not, to the best of our knowledge, been investigated systematically. This question is especially relevant when seeking a holographic description that can be compared with rotating systems studied in lattice QCD, where the plasma is formulated in a corotating reference frame \cite{lattice2021,lattice2023,lattice3,lattice4}. The importance of this distinction is further underscored by the frame dependence of the confinement--deconfinement transition temperature found in \cite{Braga2026}.

A relevant comparison is provided by recent lattice calculations of static quark--antiquark interactions in rotating QCD. The formulation of rotating lattice QCD is based on a non-inertial frame rotating with the system \cite{latticerotating1st}, and the recent quenched SU(3) lattice study of static quark--antiquark interactions adopts precisely such a rotating-frame description \cite{lattice2026}. At zero temperature, that study extracts static potentials from Wilson loops for quark pairs separated along the rotation axis and in transverse directions. At finite temperature, it uses Polyakov-loop correlators to obtain color-averaged free energies and finds position- and geometry-dependent rotational effects, with the rotational response becoming weaker as the temperature is increased \cite{lattice2026}. This provides a particularly useful first-principles reference for holographic studies of rotating heavy-quark observables. Although the lattice calculation is performed at imaginary angular velocity and in a finite rotating volume, its use of a frame corotating with the plasma makes it conceptually close to the corotating construction considered here.

These considerations indicate that the response of strongly coupled matter to rotation is not characterized by a universal behavior and that the frame in which an observable is defined can be physically significant. To address this issue in the heavy-quark sector, we study Wilson-loop observables for a quark--antiquark pair held fixed in two distinct frames. We use an equal-rotation Myers--Perry black hole in asymptotically global AdS as the holographic description of the rotating plasma. In the static frame, the quark and antiquark are fixed with respect to the static observer, and the thermodynamic temperature is obtained from the Hawking relation. In the corotating frame, the pair is fixed with respect to the rotating plasma, and the relevant local temperature is obtained from the Ehrenfest--Tolman relation. These two constructions therefore represent different physical Wilson-loop configurations rather than merely different coordinate descriptions of the same fixed quark pair. For both frames, we consider strings oriented longitudinally and transversely to the rotation and examine the effective string tension, the physical quark--antiquark separation, the maximal separation, and the heavy-quark potential. By comparing the two constructions at fixed thermodynamic temperature, we quantify how the rotational response of these Wilson-loop observables changes when the quark--antiquark pair is held fixed with respect to different observers.

The paper is organized as follows. In Sec.~2, we review the thermodynamics of the equal-rotation Myers--Perry black hole, determine the
confinement--deconfinement transition, and discuss the static and corotating temperatures. Section~3 introduces the Wilson-loop configurations and derives the effective string tensions and the expressions for the physical separation and heavy-quark potential in the
two frames. In Sec.~4, we present the numerical results for the effective string tensions and the Wilson-loop observables for longitudinal and
transverse configurations, and compare the static and corotating descriptions. Finally, Sec.~5 summarizes our main results and discusses
possible directions for future work.

%$$$$$$$$$$$$$$$$$$$$$$$$$$$$$$$$$$$$$$$$$$$$$$$$$$$$$$$$$$$$$$$$$$$$$$$$$$$$$$$$$$$$$$$$$$$$$$$$$$
\section{Rotating AdS black hole background}

We consider the five-dimensional Myers--Perry (MP) black hole, which provides a fully backreacted rotating solution of the Einstein equations with a negative cosmological constant. In five dimensions, the horizon has the topology of $S^3$ and admits two independent angular momenta associated with the two orthogonal rotation planes. We restrict to equal angular momenta, $a=b$, for which the rotational symmetry is enhanced and the geometry takes a particularly simple form \cite{MPBH}. 
This choice retains the essential features of a rotating plasma while allowing the rotation to be described by a single parameter. The conformal boundary of the geometry is $\mathbb{R}\times S^3$, so that the resulting field theory lives on a compact spatial manifold. We first describe the geometry and thermodynamics in the static frame and then introduce the frame corotating with the plasma.

%===================================================================================
\subsection{Static frame}

We use coordinates $(t,r,\theta,\psi,\phi)$, in which the conformal boundary is manifestly static and the rotation of the black hole is entirely along the $\psi$ direction \cite{MPinpsi1,MPinpsi2}. The line element of the equal-angular-momentum MP black hole is \cite{MPBH,MPinpsi1,MPinpsi2}
\begin{align}
ds^2={}& -dt^2 +\frac{dr^2}{G(r)}+\frac{r^2}{R^2}ds^2_{\rm bdy}+\frac{2\mu}{r^2}
\left[dt+\frac{a}{2}\left(d\psi+\cos\theta\,d\phi\right) \right]^2 ,
\label{eq:MP_static}
\\
ds^2_{\rm bdy} ={}&
-dt^2 +\frac{R^2}{4} \left( d\theta^2+d\psi^2+d\phi^2 +2\cos\theta\,d\psi\,d\phi \right).
\label{eq:MP_boundary_static}
\end{align}
Here $R$ is the AdS radius, $r$ is the bulk radial coordinate, and the boundary lives at $r \to \infty$ with the line element $ds^2_{\rm bdy}$. Moreover, $\mu$ and $a$ denote the effective mass and rotation parameters, respectively. The angular coordinates are chosen in the ranges $0\leq\theta\leq\pi$, $0\leq\psi<4\pi$, and $0\leq\phi<2\pi$.
The coordinate $t$ parametrizes the time direction, while the spatial part of the boundary metric describes a three-sphere of radius $R/2$ in Hopf coordinates. 

The blackening function appearing in Eq.\,\eqref{eq:MP_static} is
\begin{align}
G(r)=1+\frac{r^2}{R^2} -\frac{2\mu}{r^2} \left(1-\frac{a^2}{R^2}\right) +\frac{2a^2\mu}{r^4}.
\label{eq:G_r}
\end{align}
The outer horizon $r_+$ is the largest positive root of $G(r)$, determined by the condition $G(r_+)=0$. Solving this condition for the mass parameter gives
\begin{align}
\mu= \frac{r_+^4(R^2+r_+^2)} {2R^2r_+^2-2a^2(R^2+r_+^2)} .
\label{eq:mu_rplus}
\end{align}

The Hawking temperature, which in the static frame coincides with the thermodynamic temperature, follows from the surface gravity of the horizon and reads
\begin{align}
T_{\rm st}= \frac{
r_+^2R^2(2r_+^2+R^2) -2a^2(r_+^2+R^2)^2 }{ 2\pi r_+^2R^3 \sqrt{ r_+^2R^2-a^2(r_+^2+R^2) } }.
\label{eq:T_static}
\end{align}
The horizon is generated by the Killing vector $\xi_H=\partial_t+\Omega_{\psi} |_H\partial_\psi$. 
The choice of coordinates in Eq.\,\eqref{eq:MP_boundary_static} makes the boundary nonrotating \cite{MPinpsi1,MPinpsi2}; the angular velocity of the boundary therefore vanishes identically in both angular directions. Consequently, the angular velocity of the horizon relative to the boundary is nonzero; the horizon-generating Killing vector has a nonzero angular velocity only along $\partial_\psi$, while $\Omega_\phi|_H=0$. Since the boundary is static in these coordinates, $\Omega_\psi|_{\rm bdy}=\Omega_\phi|_{\rm bdy}=0$, and the thermodynamically relevant angular velocity entering the first law is
\begin{align}
\Omega \equiv \Omega_{\psi}|_H - \Omega_{\psi}|_{\rm bdy} = \Omega_{\psi}|_H
= -2a \left( \frac{1}{R^2}+\frac{1}{r_+^2} \right).
\label{eq:Omega_H}
\end{align}
This expression relates the rotation parameter $a$ to the physically observable angular velocity $\Omega$ of the horizon relative to the static boundary \cite{thermoMP1}. We henceforth denote this thermodynamic angular velocity simply by $\Omega$.

The thermodynamic quantities associated with the MP black hole are obtained from the gravitational action and can be expressed in terms of the horizon radius and the rotation parameter \cite{thermoMP1,thermoMP2}. The entropy, angular momentum, and total energy are, respectively,
\begin{align}
S =& \frac{4\pi^3 r_+^4 R}{\kappa^2 \sqrt{r_+^2R^2-a^2(r_+^2+R^2) }},
\label{eq:entropy_static}
\\
J=&-\frac{4a\pi^2\mu}{\kappa^2},
\label{eq:angular_momentum_static}
\\
E=& \frac{\pi\mu}{4R^2\kappa^2} \left(3R^2-a^2\right), 
\label{eq:energy_static}
\end{align}
where $\kappa^2=8\pi G_5$, with $G_5$ the five-dimensional Newton constant. The quantities in Eqs.\,\eqref{eq:entropy_static}--\eqref{eq:energy_static} satisfy the first law of black-hole thermodynamics,
\begin{align}
dE=T_{\rm st}\,dS+\Omega\,dJ.
\label{eq:first_law}
\end{align}

For the ensemble in which the temperature and angular velocity are fixed, the relevant thermodynamic potential is the grand potential $F$, obtained by the Legendre transform
\begin{align}
F=E-T_{\rm st} S-\Omega J,
\label{eq:free_energy_definition}
\end{align}
so that
\begin{align}
dF=-S\,dT_{\rm st}-J\,d\Omega.
\label{eq:free_energy_first_law}
\end{align}
At fixed angular velocity $\Omega$, this reduces to
\begin{align}
dF=-S\,dT_{\rm st},\qquad \left(\Omega=\mathrm{const.}\right).
\label{eq:dF_fixedOmega}
\end{align}
It is therefore possible to obtain the grand potential directly from the thermodynamic relation, without evaluating the Euclidean on-shell action. The additive constant is fixed by choosing pure AdS as the reference configuration, corresponding to $F\to 0$ as $r_+\to 0$. Hence,
\begin{align}
F(r_+)=-\int_{0}^{r_+}S(\tilde r_+)\frac{dT_{\rm st}(\tilde r_+)}{d\tilde r_+}\,d\tilde r_+ .
\label{eq:F_integral_zh}
\end{align}
The Hawking--Page transition is then identified by the condition
\begin{align}
F(r_{+}^c)=0.
\label{eq:HP_F_zero}
\end{align}
Substituting the entropy and temperature into Eq.\,\eqref{eq:F_integral_zh} and solving Eq.\,\eqref{eq:HP_F_zero} yields the critical horizon radius and the corresponding critical temperature,
\begin{align}
r_+^c=R \,\gamma^{1/2} , \qquad
T_c^{\rm st}(\Omega)=\frac{2 + \gamma}{2 \pi R \gamma},
\label{eq:Tcstatic}
\end{align}
where 
\begin{align}
\gamma=\frac{1}{\sqrt{1-\Omega^2R^2/4}}.
\label{eq:gamma_static}
\end{align}
The factor $R/2$ appearing here is the radius of the $\psi$-circle on the boundary $S^3$. In the absence of rotation one recovers $r_+^c=R$ and $T_c^{\rm st}(0)=T_c(0)=\frac{3}{2\pi R}$.

For the numerical analysis it is convenient to introduce the half-angular-velocity parameter
\begin{align}
\omega\equiv\frac{\Omega}{2},
\end{align}
Also, the corresponding physical tangential velocity of a fluid element on the boundary $\psi$-circle is $v=\Omega R/2 = \omega R$. Working in units $R=1$, the causality condition on the boundary requires
\begin{align}
0\leq\omega <1.
\label{eq:omega_definition}
\end{align}
The upper bound corresponds to the causal limit $v=1$ of the rotating boundary fluid. For the numerical analysis, it is convenient to introduce the radial coordinate
\begin{align}
z=\frac{R^2}{r},
\qquad
z_h=\frac{R^2}{r_+},
\label{eq:z_h_definition}
\end{align}
so that the horizon is located at $z=z_h$ and the boundary at $z=0$.

\begin{figure}[h]
\begin{center}
\includegraphics[width=7.1cm]{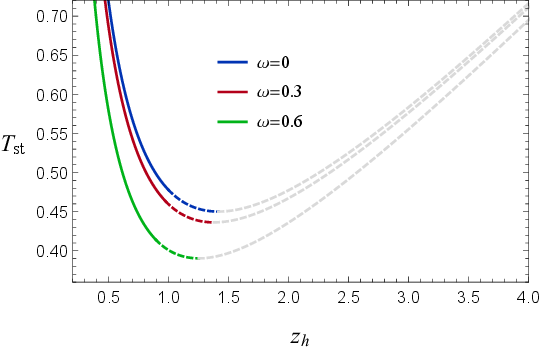}\hspace{.3cm}
\includegraphics[width=7.1cm]{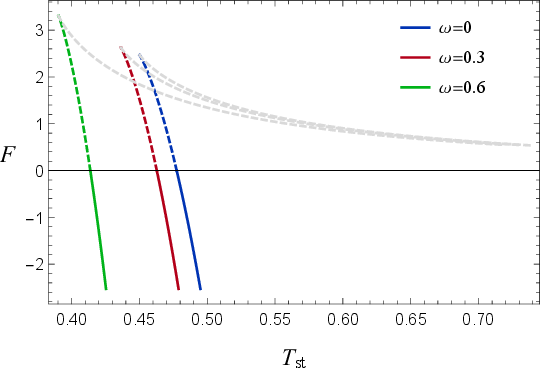}
\end{center}
\caption{\footnotesize 
Left (right): $T_{\rm st}(z_h)$ ($F(T_{\rm st})$) for various values of the angular velocity in the static frame.}
\label{TF}
\end{figure} 

The left panel of Fig.\,\ref{TF} displays the temperature as a function of the horizon position $z_h$ for $\omega=0$, $0.3$, and $0.6~{\rm GeV}$. For each value of $\omega$, the function $T_{\rm st}(z_h)$ consists of two branches separated by a minimum temperature. The branch at smaller $z_h$, corresponding to larger black holes, has positive specific heat and is locally thermodynamically stable, whereas the branch at larger $z_h$ has negative specific heat and is locally unstable. The two branches merge at the minimum of $T_{\rm st}(z_h)$, where the specific heat changes sign. This shows that, at a certain $z_h$ on the large-black-hole branch, a Hawking--Page transition between thermal AdS and the large-black-hole solution occurs, as determined from the grand potential. The grand potential associated with these black-hole solutions is shown in the right panel of Fig.\,\ref{TF}. The multivalued structure of $F(T_{\rm st})$ reflects the two branches of the black-hole solutions. The solid colored portions of the curves correspond to the thermodynamically preferred part of the large-black-hole branch, with negative grand potential $F<0$,  while the dashed colored portions denote the locally stable but thermodynamically disfavored part of the same branch with $F>0$. The unstable small-black-hole branch is shown by gray dashed curves in all cases. The Hawking--Page transition is identified by the point at which the grand potential of the large black hole vanishes, $F=0$, and hence coincides with the point where this branch crosses the temperature axis. Below the Hawking--Page temperature, thermal AdS has the lower grand potential and therefore describes the confined phase; above it, the stable large-black-hole branch has lower grand potential and corresponds to the deconfined phase. Thus, for the curves of $T_{\rm st}(z_h)$ in the left panel, the portion of the large-black-hole branch above the Hawking--Page temperature is thermodynamically preferred, whereas its continuation below the transition remains locally stable but is thermodynamically disfavored relative to thermal AdS.

\begin{figure}[h]
\begin{center}
\includegraphics[width=7.4cm]{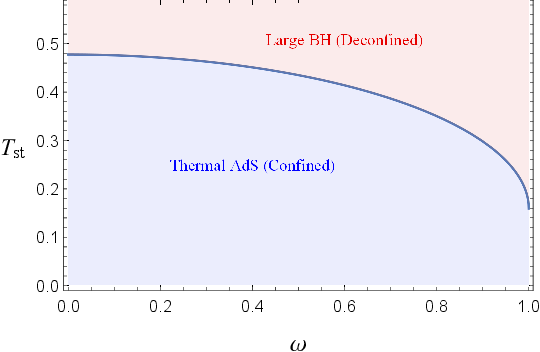}
\end{center}
\caption{\footnotesize 
The Hawking--Page transition temperature versus the angular velocity $\omega$ in the static frame.}
\label{Tcs}
\end{figure} 

Increasing the angular velocity lowers the temperature at fixed $z_h$, as seen in the left panel of Fig.\,\ref{TF}. At the same time, the zero of the grand potential shifts toward lower temperatures as $\omega$ increases, in agreement with the analytical result in Eq.\,\eqref{eq:Tcstatic}. Consequently,  the Hawking--Page temperature measured in the static frame decreases monotonically with increasing angular velocity, as shown in Fig.\,\ref{Tcs}. The corresponding temperature measured in the corotating frame will be discussed in the next subsection.

%==================================================================================
\subsection{Corotating frame}

To study the response of a quark--antiquark pair in a rotating plasma, it is useful to work in a frame that corotates with the plasma. In this frame the plasma is locally at rest, and a quark pair corotating with the plasma remains at fixed spatial coordinates. The coordinate transformation from the static frame to the corotating frame is
\begin{align}
\psi \rightarrow \psi+\Omega t ,
 \label{eq:corotating_transformation}
\end{align}
where $\psi$ on the right-hand side denotes the angular coordinate measured by the corotating observer, and $\Omega$ is the thermodynamic angular velocity defined in Eq.\,\eqref{eq:Omega_H}. For convenience we retain $\omega=\Omega/2$, so that the Lorentz factor associated with the corotating motion at the boundary radius $R/2$ is
\begin{align}
\gamma= \frac{1}{\sqrt{1-\omega^2R^2}},
\label{eq:gamma_definition}
\end{align}
which coincides with the quantity defined in Eq.\,\eqref{eq:gamma_static}.

Applying the transformation \eqref{eq:corotating_transformation} to the bulk and boundary metrics in Eqs.\,\eqref{eq:MP_static} and \eqref{eq:MP_boundary_static}, and switching to the radial coordinate $z$, the corotating bulk and boundary metrics become
\begin{align}
ds^2_{\rm cor}={}& -dt^2 +\frac{R^4}{z^4G(z)}\,dz^2 +\frac{R^2}{z^2}\,ds^2_{\rm bdy,cor}%\nonumber\\&
+\frac{2\mu z^2}{R^4} \left[ \left(1+a\,\omega \right)dt +\frac{a}{2} \left(d\psi+\cos\theta\,d\phi \right) \right]^2 ,
\label{eq:MP_corotating_compact}\\
ds^2_{\rm bdy,cor}={}&-\frac{dt^2}{\gamma^2}+\frac{R^2}{4}\left(d\theta^2+d\psi^2+d\phi^2+2\cos\theta\,d\psi\,d\phi\right)
+R^2\omega \,dt \left(d\psi+\cos\theta\,d\phi \right) .
\label{eq:boundary_corotating}
\end{align}

The temperature relevant for a corotating observer differs from the Hawking temperature defined with respect to the static-frame time coordinate. In thermal equilibrium, the local temperature measured by a comoving observer is related to the global temperature by the Ehrenfest--Tolman redshift law, which in our case takes the form $T\sqrt{-g_{tt}}=\mathrm{const}$ along any stationary flow \cite{Tolman1930}. Evaluating $g_{tt}$ at the conformal boundary in the corotating and static frames yields the redshift factor $\gamma$, so that the local temperature measured by the corotating observer is \cite{Braga2026}
\begin{align}
T_{\rm cor}=\gamma\, T_{\rm st},
\label{eq:T_cor}
\end{align}
where $\gamma$ is given by Eq.\,\eqref{eq:gamma_definition} and $T_{\rm st}$ is the static-frame Hawking temperature, Eq.\,\eqref{eq:T_static}. Rotation therefore increases the temperature seen by the corotating observer at fixed angular velocity. In particular, the Hawking--Page transition temperature in the corotating frame is
\begin{align}
T_c^{\rm cor}(\Omega)=\gamma\,T_c^{\rm st}(\Omega).
\label{eq:Tc_cor}
\end{align}
Combining Eq.\,\eqref{eq:Tcstatic} with the Tolman factor, the critical temperature measured by the corotating observer becomes
\begin{equation}
T_c^{\rm cor}(\Omega)=\frac{2+\gamma}{2\pi R}.
\label{eq:Tc-cor-final}
\end{equation}
At $\Omega=0$ we recover $\gamma=1$ and $T_c^{\rm cor}(0)=T_c(0)=3/(2\pi R)$, as it should be. For $\Omega\neq0$, however, $\gamma>1$ and Eq.\,\eqref{eq:Tc-cor-final} shows that $T_c^{\rm cor}$ \emph{increases} with rotation monotonically, in contrast to $T_c^{\rm st}(\Omega)$ in the static frame, which decreases. The two temperatures differ only by the local blueshift factor $\gamma$ between the static and the corotating boundary observer, yet this factor is enough to flip the sign of the effect.

This reversal is not a small correction: it changes the qualitative prediction of the model. A static-frame observer would conclude, as in most of the earlier holographic constructions including genuine, boosted or globally rotating backgrounds, that rotation makes deconfinement easier. A corotating observer, i.e., the frame in which the plasma itself, and in principle a comoving lattice simulation, would measure the transition, concludes the opposite. Since the plasma itself carries the angular momentum in heavy-ion collisions, the corotating temperature is the appropriate quantity for comparison with such measurements.

It is then natural to ask whether other observables sensitive to the heavy-quark sector exhibit a similar dependence on the frame in which the rotating plasma is characterized. Wilson loops provide a natural framework for addressing this question, since they allow us to study both the effective string tension and the heavy-quark potential in the rotating geometry. Previous holographic studies have found different rotational effects on these quantities, depending on the holographic realization of rotation and on the orientation of the quark--antiquark pair. In particular, while most constructions find a suppression of the heavy-quark potential with increasing angular velocity, the Kerr--AdS description considered in \cite{KerrAdSwilson1} yields an increasing heavy-quark potential in the longitudinal configuration, together with a decreasing interquark distance. It is therefore interesting to investigate how the heavy-quark sector is characterized when the quark--antiquark pair is taken to be static in the corotating frame. In the next section, we construct the Wilson loop directly in the corotating metric of Eqs.\,\eqref{eq:MP_corotating_compact} and \eqref{eq:boundary_corotating}, and study the effective string tension, quark separation, and heavy-quark potential for longitudinal and transverse configurations. We then compare these results with the corresponding Wilson-loop observables calculated in the static frame, where the quark--antiquark pair is static with respect to the nonrotating observer. This comparison allows us to  examine how the heavy-quark sector is characterized when the quarks are taken to be comoving with the rotating plasma rather than static in the nonrotating frame.

%**************************************************************************************************************************
\section{Wilson loop and heavy quark potential}

\subsection{Heavy quark potential from the Wilson loop in a general setup}

The potential between a static quark--antiquark pair separated by a distance $\Delta q$ along a boundary direction $q$ can be extracted from the expectation value of a rectangular Wilson loop,
\begin{align}
V(\Delta q)=-\lim_{\mathcal{T}\to\infty}\frac{1}{\mathcal{T}} \ln\langle W[\mathcal{C}]\rangle ,
\label{eq:Wilson_loop_potential}
\end{align}
where $\mathcal{T}$ denotes the temporal extent of the loop. In the strong-coupling limit, the Wilson loop is described holographically by the classical Nambu--Goto action of a fundamental string whose endpoints are attached to the quark and antiquark at the boundary. We consider a stationary five-dimensional bulk geometry and write the explicit form of part of the metric relevant for a string stretched along a spatial direction $q$ as
\begin{align}
ds^2={}&g_{tt}(z)\,dt^2+2g_{tq}(z)\,dt\,dq+g_{qq}(z)\,dq^2+g_{zz}(z)\,dz^2+g_{\perp}(z)\,d\vec{x}_{\perp}^{\,2},
\label{eq:general_metric_WL}
\end{align}
where the remaining spatial coordinates are collectively denoted by $\vec{x}_{\perp}$. We assume that the metric functions depend only on the holographic coordinate $z$, so that the geometry is stationary and invariant under translations along the chosen direction $q$. This form will allow us to treat the longitudinal and transverse orientations in our rotating background within the same framework.

The Nambu--Goto action is
\begin{align}
S_{\rm NG}=T_f\int d\tau\,d\xi\,\sqrt{-\det h_{\alpha\beta}},
\qquad
T_f=\frac{1}{2\pi\alpha'},
\label{eq:NG_action_general}
\end{align}
where $T_f$ is the fundamental string tension and $h_{\alpha\beta}$ is the induced metric on the string worldsheet. For a quark--antiquark pair aligned along the $q$ direction, we choose the static gauge
\begin{align}
t=\tau, \qquad q=\xi, \qquad z=z(\xi),
\label{eq:string_embedding_general}
\end{align}
while all other bulk coordinates are kept fixed. The nonvanishing components of the induced metric are then
\begin{align}
h_{\tau\tau}&=g_{tt},\quad h_{\tau\xi}=g_{tq}, \quad h_{\xi\xi}=g_{qq}+g_{zz}z'(q)^2,
\label{eq:induced_metric_general}
\end{align}
where the prime denotes differentiation with respect to $q$. The Nambu--Goto action consequently takes the form
\begin{align}
S_{\rm NG}=T_f\mathcal{T} \int dq\, \sqrt{A(z)z'(q)^2+\sigma^2(z)},
\label{eq:NGgeneral}
\end{align}
where we have introduced
\begin{align}
A(z) &\equiv -g_{tt}(z)g_{zz}(z),
\label{eq:A_general}
\\
\sigma(z) &\equiv \sqrt{ g_{tq}^2(z)-g_{tt}(z)g_{qq}(z) }.
\label{eq:sigma_general}
\end{align}
The quantity $\sigma(z)$ is the metric combination that controls the energy of a horizontal string segment in the $q$ direction. Accordingly, the corresponding effective string tension is
\begin{align}
\sigma_{\rm eff}(z) = T_f\sigma(z) = T_f \sqrt{ g_{tq}^2(z)-g_{tt}(z)g_{qq}(z) }.
\label{eq:sigma_eff_general}
\end{align}
For a metric without a $tq$ cross term, this reduces to the familiar form $\sigma_{\rm eff}=T_f\sqrt{-g_{tt}g_{qq}}$. In the present problem, the additional $g_{tq}$ contribution is essential and will distinguish the longitudinal configuration from the transverse one.

Since the Lagrangian in Eq.\,\eqref{eq:NGgeneral} has no explicit dependence on $q$, the corresponding Hamiltonian is conserved. It is given by
\begin{align}
{\cal H} &= z'\frac{\partial\mathcal{L}}{\partial z'} -\mathcal{L} = -\frac{\sigma^2(z)} {\sqrt{\sigma^2(z)+A(z)z'^2}},
\end{align}
where the overall sign is irrelevant for the conserved quantity. 
At the turning point of the connected string, denoted by $z=z_c$, the profile satisfies $z'(q)\big|_{z=z_c}=0$ and hence 
${\cal H}_c=\sigma(z_c)$. 
Solving ${\cal H}={\cal H}_c$ for the string profile gives $dq$. Then, integrating from the boundary to the turning point gives the coordinate separation of the two endpoints 
\begin{align}
\Delta q=2\int_0^{z_c}dz\,\frac{\sigma(z_c)\sqrt{A(z)}}{\sigma(z)\sqrt{\sigma^2(z)-\sigma^2(z_c)}}.
\label{eq:Deltaq}
\end{align}
The factor of two follows from the reflection symmetry of the connected U-shaped string about its turning point.

The on-shell action of the connected configuration is ultraviolet divergent because the endpoints of the string reach the AdS boundary, corresponding to the infinite bare masses of the external quark and antiquark. We remove this divergence by subtracting the self-energy of the two isolated quarks, represented by two straight strings extending from the boundary to the horizon at fixed $q$. For such a configuration, the induced metric contains only the $t$ and $z$ directions, and its contribution to the energy of each string is $T_f\int_0^{z_h} dz\,\sqrt{A(z)}$. The renormalized heavy-quark potential is therefore
\begin{align}
V(\Delta q)=2T_f\left\{\int_0^{z_c}dz\,\sqrt{A(z)}\left[\frac{\sigma(z)}{\sqrt{\sigma^2(z)-\sigma^2(z_c)}}-1\right]-\int_{z_c}^{z_h}dz\,\sqrt{A(z)}\right\}.
\label{eq:Vgeneral}
\end{align}
Equations \eqref{eq:Deltaq} and \eqref{eq:Vgeneral} provide the general relations needed to determine the heavy-quark separation
and potential once the appropriate metric components are specified.

In the rotating geometry considered below, the relevant choice of the boundary direction $q$ depends on the orientation of the quark--antiquark pair relative to the rotation. The rotational symmetry of the equal-angular-momenta Myers--Perry solution allows us to identify a longitudinal configuration associated with the rotational direction and a transverse configuration orthogonal to it. We discuss these two embeddings and the corresponding physical boundary separation in the next subsection.

%************************************************************************************************************************
\subsection{Longitudinal and transverse configurations }

We now specialize the general Wilson-loop construction to the rotating background introduced in Sec.2. In the equal-rotation Myers--Perry geometry considered here,  the angular momentum is associated with the Killing direction $\partial_\psi$ and the horizon generator is characterized by a nonvanishing angular velocity $\Omega_\psi$, whereas the angular velocity associated with the second azimuthal Killing direction vanishes, $\Omega_\phi=0$.
Therefore, the $\psi$ direction provides the natural longitudinal orientation with respect to the rotation. We consider a quark--antiquark pair at the boundary that is static in the corotating frame, with its separation oriented along the $\psi$ direction. The corresponding string embedding is chosen as
\begin{align}
t=\tau,\quad \psi=\xi, \quad z=z(\xi), \quad \theta=\theta_0, \quad \phi=\phi_0,
\label{eq:long_embedding}
\end{align}
where $\theta_0$ and $\phi_0$ specify the location of the pair on the remaining directions of the boundary $S^3$ and $\xi$ parametrizes the worldsheet.

It is important that the nonvanishing $g_{t\psi}$ component in the corotating frame does not invalidate this choice of orientation. Rather, such a time--angular mixing is a characteristic feature of a rotating, stationary but non-static spacetime. Since the angular momentum of the solution is associated with the Killing direction $\partial_\psi$, this direction provides the natural longitudinal direction for the Wilson loop. The effect of the rotation, including the frame-dragging contribution encoded in $g_{t\psi}$, then enters directly through the induced worldsheet metric. In particular, for the embedding in Eq.\,\eqref{eq:long_embedding}, the nonvanishing components of the induced metric are
\begin{align}
  h_{\tau\tau} = g_{tt}, \quad
  h_{\tau\xi} = g_{t\psi}, \quad
  h_{\xi\xi} = g_{\psi\psi} + g_{zz}\,z'^2 ,
\label{eq:induced_metric_long}
\end{align}
Consequently, the effective string tension that determines the horizontal part of the string energy is
\begin{align}
\sigma_{\rm eff}^{\parallel}(z)=T_f \sigma_{\parallel}(z)=T_f\sqrt{ g_{t\psi}^2-g_{tt}\,g_{\psi\psi} }.
\label{eq:sigma_eff_L}
\end{align}

The physical separation of the quark and antiquark is determined by the boundary spatial geometry. Since the two endpoints are fixed with respect to the corotating observer, their separation is measured on a constant-$t$ hypersurface. Restricting the boundary line element to the longitudinal direction, with $dt=d\theta=d\phi=0$, gives
\begin{align}
dL_{\parallel}^2= \frac{R^2}{4}\,d\psi^2 .
\label{eq:boundary_dl_L}
\end{align}
Hence the angular-coordinate separation $\Delta\psi$ obtained from the string profile by integrating Eq.\,\eqref{eq:Deltaq} is converted to the physical longitudinal separation according to
\begin{align}
L_{\parallel} = \frac{R}{2}\Delta\psi .
\label{eq:L_L}
\end{align}

For the transverse configuration, we choose the $\theta$ direction. This is a particularly simple choice because the spatial metric contains no $d\theta\,d\psi$ or $d\theta\,d\phi$ cross terms. Thus, a curve with $\psi$ and $\phi$ fixed and varying $\theta$ is orthogonal to the longitudinal $\psi$ direction. The corresponding string embedding is
\begin{align}
t=\tau, \quad \theta=\xi, \quad z=z(\xi), \quad \psi=\psi_0, \quad \phi=\phi_0,
\label{eq:trans_embedding}
\end{align}
and the relevant induced metric components are
\begin{align}
h_{\tau\tau}=g_{tt},\quad
h_{\tau\xi}=0,\quad
h_{\xi\xi}=g_{\theta\theta}+g_{zz}z'^2 .
\label{eq:induced_metric_transverse}
\end{align}
Here $g_{t\theta} = 0$, and the corresponding effective string tension therefore reduces to
\begin{align}
\sigma_{\rm eff}^{\perp}(z) =T_f \sigma_{\perp}(z) = T_f\sqrt{-g_{tt}\,g_{\theta\theta}} .
\label{eq:sigma_eff_T}
\end{align}
On a constant-$t$ boundary slice, the transverse line element for the chosen $\theta$ direction is
\begin{align}
dL_{\perp}^2=\frac{R^2}{4}\,d\theta^2 ,
\label{eq:boundary_dl_T}
\end{align}
and therefore the physical transverse separation is
\begin{align}
L_{\perp}=\frac{R}{2}\Delta\theta .
\label{eq:L_T}
\end{align}
Thus, in both orientations, the angular coordinate used to parametrize the string is a dimensionless worldsheet coordinate, while the physical quark separation is obtained only after converting the boundary coordinate separation using the boundary metric.

The other azimuthal coordinate $\phi$, although independent of the rotational Killing coordinate $\psi$, does not in general provide a direction transverse to the rotation. At a generic value of $\theta$, the boundary metric contains the cross term
\begin{align}
g_{\psi\phi} = \left( \frac{R^4}{4z^2} +\frac{\mu a^2z^2}{2R^4} \right)\cos\theta ,
\label{eq:psi_phi_cross}
\end{align}
which vanishes only at $\theta=\pi/2$. Thus, a curve obtained by varying $\phi$ while keeping $\psi$ fixed is not orthogonal to the
rotational direction, except on the equator. More generally, an explicitly orthogonal direction in the $(\psi,\phi)$ sector can be
constructed as $ \partial_\phi-\cos\theta\,\partial_\psi $, at arbitrary $\theta$. We do not need this more general choice here. Instead, we choose the $\theta$ direction, which is orthogonal to $\partial_\psi$ everywhere and therefore provides a globally well-defined transverse configuration that is also technically simpler for the Wilson-loop calculation.

It is useful to note how the effective string tensions transform under the change from the static to the corotating coordinates. Under the transformation $\psi\rightarrow\psi+\Omega t$, the relevant metric components transform as
\begin{align}
g_{tt}^{\rm cor}&=g_{tt}^{\rm st} +2\Omega g_{t\psi}^{\rm st} +\Omega^2 g_{\psi\psi},
\nonumber\\
g_{t\psi}^{\rm cor}&=g_{t\psi}^{\rm st} +\Omega g_{\psi\psi},
\label{eq:metric_transform_tpsi}
\end{align}
while $g_{\psi\psi}$ and $g_{zz}$ remain unchanged. Consequently,
\begin{align}
\left(g_{t\psi}^{\rm cor}\right)^2 -g_{tt}^{\rm cor}g_{\psi\psi} = \left(g_{t\psi}^{\rm st}\right)^2 -g_{tt}^{\rm st}g_{\psi\psi},
\label{eq:sigma_L_invariant}
\end{align}
so that the longitudinal effective string tension is unchanged by the coordinate transformation when evaluated at the same bulk geometry, or equivalently at the same horizon position $z_h$,
\begin{align}
\sigma_{\rm eff}^{\parallel,\,\rm cor}(z;z_h) = \sigma_{\rm eff}^{\parallel,\,\rm st}(z;z_h).
\end{align}
This does not imply that the Wilson-loop observables are identical when the two frames are compared at the same physical temperature. Moreover, the quantity $A(z)=-g_{tt}(z)g_{zz}(z)$ entering the separation and potential is frame dependent even at the level of the bulk geometry, because $g_{tt}$ changes under the transformation. Although the longitudinal effective string tension is invariant under the coordinate transformation at fixed $z_h$, the transverse effective string tension is likewise frame dependent even at the same $z_h$,
\begin{align}
\sigma_{\rm eff}^{\perp,\,\rm cor} =T_f\sqrt{-g_{tt}^{\rm cor}\,g_{\theta\theta}},\quad
\sigma_{\rm eff}^{\perp,\,\rm st} =T_f\sqrt{-g_{tt}^{\rm st}\,g_{\theta\theta}}.
\label{eq:sigma_T_frames}
\end{align}
The separation lengths and heavy-quark potentials must also be evaluated separately in the two frames, since the Wilson loops describe
quark--antiquark pairs held fixed with respect to different observers.

The general relations of Sec.~3.1 can now be applied directly to the two embeddings by using the appropriate effective string tension and metric functions for each orientation. Substituting Eqs.\,\eqref{eq:sigma_eff_L} and \eqref{eq:sigma_eff_T} into the general expressions \eqref{eq:Deltaq} and \eqref{eq:Vgeneral}, and converting the coordinate separations according to Eqs.\,\eqref{eq:L_L} and \eqref{eq:L_T}, we obtain the physical separations $L_\parallel$ and $L_\perp$ and the corresponding heavy-quark potentials $V(L_\parallel)$ and $V(L_\perp)$. In the two descriptions, these quantities are functions of the angular velocity $\Omega$ and the corresponding temperature, $T_{\rm cor}$ for the corotating frame and $T_{\rm st}$ for the static frame. In Sec.~4, we present the resulting effective string tensions, separation lengths, and heavy-quark potentials in the corotating frame, and then compare them with the corresponding Wilson-loop observables calculated in the static frame, where the quark--antiquark pair is static with respect to the nonrotating observer. This comparison allows us to examine how the heavy-quark sector is characterized when the quarks are taken to be comoving with the rotating plasma rather than static in the nonrotating frame.

%*************************************************************************************************************************
\section{Results}

We now investigate the effect of rotation on the effective string tension, the maximum separation of a connected quark--antiquark pair, and the heavy-quark potential in the longitudinal and transverse directions. The calculations in the main part of this section are performed for quark--antiquark pairs that are static in the corotating frame, at fixed corotating temperature $T_{\rm cor}$. For each orientation, the connected string configurations are obtained from the relations derived in Sec.~3. We first discuss the effective string tension and then consider the separation length and the heavy-quark potential for the two orientations separately. We subsequently compare these quantities with the corresponding quantities calculated in the static frame of the same rotating geometry. In the numerical analysis below, we set $R=1~{\rm GeV}^{-1}$ without loss of generality. Unless otherwise stated, the dimensionful quantities appearing in the figures are given in their natural units: $T_{\rm cor}$, $T_{\rm st}$, $\omega$, and $V$ are in ${\rm GeV}$, while $z$, $z_c$, and the quark separation lengths are in ${\rm GeV}^{-1}$. The units are omitted from the axis labels and legends for brevity.
%=========================================================================
\subsection{Effective string tension}

The effective string tension provides a direct measure of the energy cost of stretching a string in a given spatial direction. In the present setup, the longitudinal and transverse tensions in the corotating frame are obtained from Eqs.\,\eqref{eq:sigma_eff_L} and \eqref{eq:sigma_eff_T}, respectively. Since their absolute values vary substantially with the radial position, it is useful to consider the ratio
\begin{align}
\widehat{\sigma}_{\rm eff}^{\,i}(z,z_h(T_{\rm cor},\omega),\omega) = \frac{ \sigma_{\rm eff}^{\,i} \bigl(z,z_h(T_{\rm cor},\omega),\omega\bigr) }{
\sigma_{\rm eff}^{\,i} \bigl(z,z_h(T_{\rm cor},0),0\bigr) },
\qquad i=\parallel,\perp ,
\label{eq:normalized_sigma_results}
\end{align}
which equals unity at $\omega=0$ for every value of $z$.  This normalization isolates the rotational modification of the tension from
its overall radial dependence and allows the rotational modification at different bulk positions to be compared on the same scale.

\begin{figure}[h]
\begin{center}
\includegraphics[width=7.1cm]{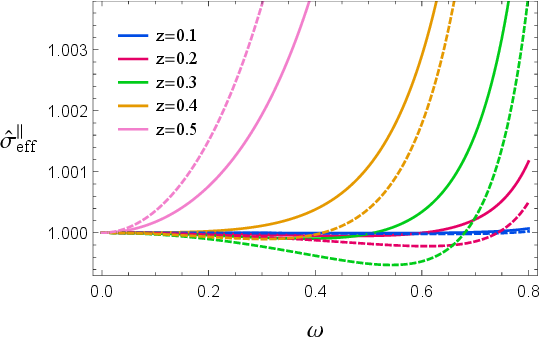}\hspace{.3cm}
\includegraphics[width=7.1cm]{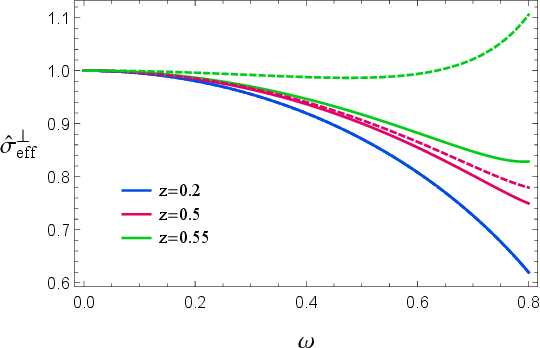}
\end{center}
\caption{\footnotesize 
Normalized effective string tension as a function of the angular velocity $\omega$ for several values of the bulk radial coordinate $z$, shown for the longitudinal (left) and transverse (right) directions in the corotating frame. Solid and dashed curves correspond to $T_{\rm cor}=0.6~{\rm GeV}$ and $0.65~{\rm GeV}$, respectively.}
\label{sigmacor}
\end{figure}

Figure \ref{sigmacor} shows the normalized effective string tensions in the corotating frame for several values of $z$ and two values of $T_{\rm cor}$. Solid (dashed) lines represent the results for $T_{\rm cor}=0.6~{\rm GeV}$ ($T_{\rm cor}=0.65~{\rm GeV}$). As mentioned earlier, the transition temperature to the deconfined phase increases with the angular velocity in the corotating frame. Therefore, for a chosen $T_{\rm cor}$, $\omega$ cannot be increased arbitrarily up to the causal limit of the rotating fluid, $\omega=1$. In all the results presented here and below, we restrict $\omega$ to values for which the system remains in the deconfined phase, so that a black-hole solution exists and the calculations based on the metric in Sec.~2.2 remain applicable. Moreover, the values of $z$ are chosen such that they remain smaller than $z_h$ for the corresponding black-hole solutions.

The longitudinal and transverse directions exhibit qualitatively different responses to rotation. In the longitudinal case, the rotational correction is relatively small at small values of $z$. There, the tension initially decreases with increasing $\omega$ and subsequently turns to an increasing behavior at larger angular velocities. As one moves deeper into the bulk, the rotational correction becomes substantially larger, and at sufficiently large $z$ the longitudinal tension increases monotonically with $\omega$ over the whole range considered. The transverse tension generally decreases with increasing $\omega$ at lower values of $z$ and $\omega$. An increasing behavior appears only at sufficiently large $z$ and sufficiently large $\omega$, with the onset of this behavior occurring at smaller values of $z$ for higher $T_{\rm cor}$. Thus, at sufficiently large radial positions, the rotational modifications of the longitudinal and transverse tensions can become comparable in magnitude, although their behavior at smaller $z$ is distinct. The results therefore do not indicate a universal enhancement or suppression of the effective string tension by rotation; rather, its response depends on both the orientation of the string and its radial position in the bulk.

The comparison between the two temperatures also shows that the effect of temperature on the transverse tension, relative to its rotational dependence, is less pronounced than for the longitudinal tension, particularly at lower values of $z$. As can be seen in the right panel, the transverse tensions for the two temperatures considered here cannot be distinguished on the scale of the rotational variation at $z=0.2~{\rm GeV}^{-1}$, and the curves for this value of $z$ at $T_{\rm cor}=0.6~{\rm GeV}$ and $0.65~{\rm GeV}$ nearly overlap over the displayed range of $\omega$.

\begin{figure}[h]
\begin{center}
\includegraphics[width=7.1cm]{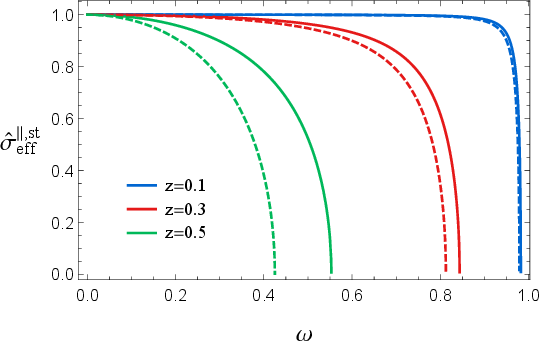}\hspace{.3cm}
\includegraphics[width=7.1cm]{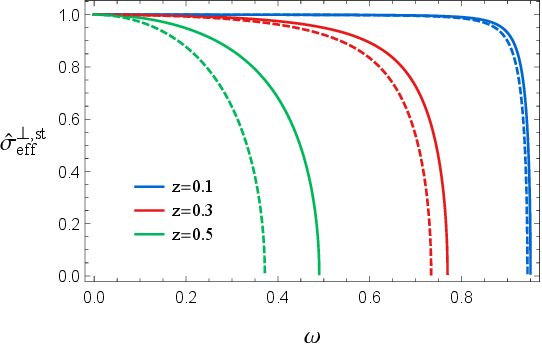}
\end{center}
\caption{\footnotesize
Normalized effective string tension as a function of the angular velocity $\omega$ in the static frame for several values of the bulk radial coordinate $z$. The longitudinal and transverse orientations are shown in the left and right panels, respectively. Solid and dashed curves correspond to $T_{\rm st}=0.6~{\rm GeV}$ and $0.65~{\rm GeV}$, respectively.}
\label{sigma-st}
\end{figure}

It is useful to contrast these results with the effective string tensions obtained when the same rotating geometry is described in the static frame. In this description, the Wilson loop is constructed for a quark--antiquark pair that is static with respect to the static observer and therefore represents a physical configuration different from the one considered in the corotating frame. Figure~\ref{sigma-st} shows the normalized effective string tensions as function of $\omega$ in the static frame for the longitudinal (left) and transverse (right) orientations, for $T_{\rm st}=0.6$ and $0.65~{\rm GeV}$ and several values of the bulk radial coordinate.

For the longitudinal configuration, the effective string tension in the static frame decreases monotonically with increasing $\omega$ at fixed $z$ and $T_{\rm st}$, in contrast to the increasing behavior that can arise in the corotating frame at sufficiently large $z$ and $\omega$. The difference is not in the algebraic form of the longitudinal tension itself. At fixed values of the geometric parameters, the combination
\begin{align}
\sigma_{\rm eff}^{\parallel}=T_f\sqrt{g_{t\psi}^2-g_{tt}g_{\psi\psi}}
\end{align}
has the same dependence on the rotation parameter in the two coordinate descriptions. However, the quantity considered in the numerical analysis is the normalized $\sigma_{\rm eff}^{\parallel}(z,T,\omega)$ at fixed thermodynamic temperature $T$. Since the relation between the horizon position and the temperature is different in the two frames, fixing $T_{\rm st}$ rather than $T_{\rm cor}$ changes the resulting $\omega$ dependence. 

As can be seen in Fig.\,\ref{sigma-st}, both static-frame tensions decrease with $\omega$ throughout the displayed range and eventually vanish at a finite value of the angular velocity within the causally allowed range. Moreover, increasing either $T_{\rm st}$ or $z$ enhances the suppression by rotation: the tensions become smaller and reach zero at a lower value of $\omega$. Moreover, the decrease of the transverse tension with rotation is more than the longitudinal tension, at fixed $T_{\rm st}$ and $z$. Thus, within the static-frame description, no increasing behavior of either longitudinal or transverse tensions is found in the radial range accessible to the black-hole geometry. However, the vanishing of the longitudinal tension has a different geometric origin from that of the transverse tension. For the longitudinal configuration, the effective string tension vanishes at the event horizon. Consequently, for a fixed $z$ and $T_{\rm st}$, the zero of the longitudinal tension as a function of $\omega$ occurs when the corresponding event horizon reaches that value of $z$. This should be distinguished from the transverse configuration, for which the zero occurs at the outer surface $z_*$ where $g_{tt}^{\rm st}=0$. Since $g_{t\theta}=0$,
\begin{align}
\sigma_{\rm eff}^{\perp,{\rm st}}=T_f\sqrt{-g_{tt}^{\rm st}\,g_{\theta\theta}},
\end{align}
and hence $\sigma_{\rm eff}^{\perp,{\rm st}}$ vanishes when $g_{tt}^{\rm st}$ vanishes. For the rotating geometries considered here, this surface lies outside the event horizon, $z_*<z_h$. Thus, the zero of the transverse tension signals the presence of a worldsheet horizon in the static-frame geometry, in analogy with the worldsheet horizons that arise for strings describing heavy quarks moving through a thermal plasma \cite{zstar1,zstar2,zstar3,zstar4,zstar5}. The corresponding feature appears explicitly in the transverse separation when the turning point approaches $z_*$, as discussed in Sec.~4.3.

The comparison with the corotating results is consequently more subtle than a comparison of the explicit metric expressions alone. In the longitudinal direction, the same metric combination determines the tension in both frames, but evaluating it at fixed thermodynamic temperature leads to different $\omega$ dependences because the horizon--temperature relation differs between the two descriptions. In the transverse direction, the tension already differs at the level of the metric through the different $g_{tt}$ components. The static-frame calculation therefore provides a useful comparison with the corotating-frame result while emphasizing that the two Wilson loops correspond to quark--antiquark pairs that are static in different frames. In the following subsections, we examine how these frame-dependent differences in the effective tension are reflected in the boundary separation and the heavy-quark potential.

%=========================================================================
\subsection{Longitudinal Wilson loop and heavy-quark potential}

We first consider a quark--antiquark pair separated along the longitudinal direction, identified with the $\psi$ direction of the rotating geometry. As discussed in Sec.~3, the angular separation of the endpoints obtained from the string profile is converted to the physical boundary separation using the spatial boundary metric on a constant-$t$ slice. The resulting quantity is therefore the physical separation $L_\parallel$, rather than the coordinate difference $\Delta\psi$.

\begin{figure}[h]
\begin{center}
\includegraphics[width=7.1cm]{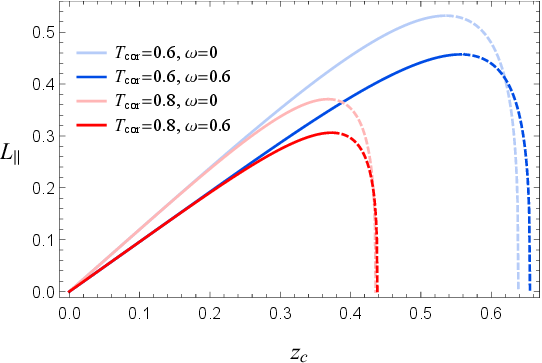}\hspace{.3cm}
\includegraphics[width=7.1cm]{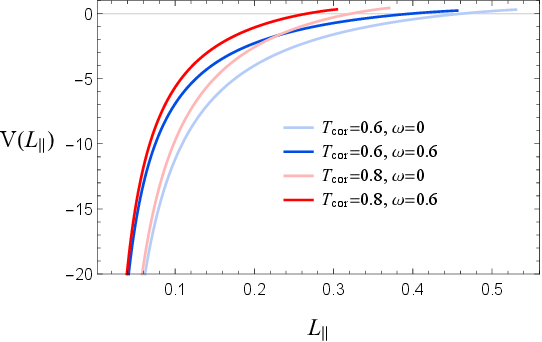}
\end{center}
\caption{\footnotesize
The quark-antiquark separation $L_\parallel$ as a function of the turning point $z_c$ (left) and the corresponding heavy-quark potential $V(L_\parallel)$ (right) in the corotating frame for two values of $T_{\rm cor}$ and for $\omega=0$ and $\omega=0.6~{\rm GeV}$.}
\label{LV-L}
\end{figure}

The left panel of Fig.\,\ref{LV-L} shows $L_\parallel$ as a function of the turning point of the U-shaped string configuration, $z_c$, for $T_{\rm cor}=0.6$ and $0.8~{\rm GeV}$, with and without rotation. For a fixed temperature, the connected solution exhibits the usual two-branch structure corresponding to two U-shaped string configurations in the deconfined phase. The branch with smaller $z_c$, corresponding to a string closer to the boundary, is energetically preferred, whereas the continuation to larger $z_c$ is locally stable but thermodynamically disfavored. The two branches merge at a maximal separation, beyond which no connected U-shaped solution exists. The physical configuration beyond this point is then described by two disconnected straight strings extending from the boundary to the event horizon.

\begin{figure}[h]
\begin{center}
\includegraphics[width=7.1cm]{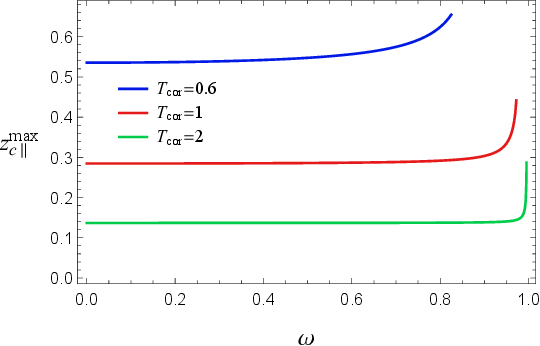}\hspace{.3cm}
\includegraphics[width=7.1cm]{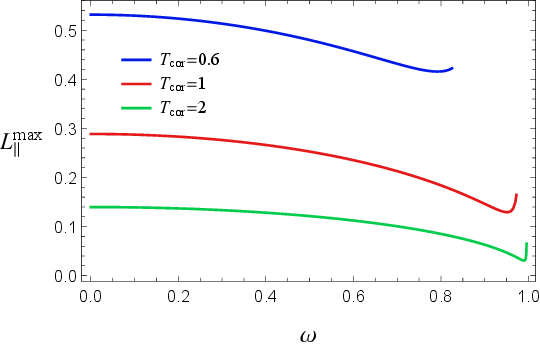}
\end{center}
\caption{\footnotesize
The turning point $z_{c,\parallel}^{\rm max}$ corresponding to the maximum longitudinal separation (left) and the maximum separation $L_\parallel^{\rm max}$ (right) as functions of the angular velocity $\omega$ in the corotating frame for three values of $T_{\rm cor}$.}
\label{maxL}
\end{figure}

Rotation modifies this structure in a definite way. Increasing $\omega$ shifts the longitudinal separation $L_\parallel$ to smaller values at fixed $z_c$ and consequently reduces the maximal separation $L_\parallel^{\rm max}$. This effect is already visible in the $L_\parallel(z_c)$ curves and becomes more transparent in the corresponding $L_\parallel^{\rm max}(\omega)$ plot in the right panel of Fig.\,\ref{maxL}. The decrease of the maximal separation indicates that rotation makes the connected longitudinal configuration unable to sustain as large a boundary separation as in the nonrotating case.

The corresponding heavy-quark potentials are shown in the right panel of Fig.\,\ref{LV-L}. At both temperatures considered here, increasing $\omega$ raises the potential at fixed separation. The connected potential is negative at sufficiently small separation, where the connected string configuration is energetically favored. As the separation increases, the potential approaches the energy of the disconnected configuration. In the present subtraction scheme adopted here, the zero of the renormalized potential identifies the separation at which the connected and disconnected configurations have equal energy. Beyond this point, whenever the connected branch persists, the disconnected configuration is energetically preferred.

The behavior of the heavy-quark potential provides information that is distinct from that contained in the maximal separation. While
$L_\parallel^{\rm max}$ determines the range of boundary separations for which a connected string solution exists, $V(L_\parallel)$ characterizes the energetic competition between the connected and disconnected configurations within this range. Rotation therefore affects both the existence of connected configurations and their energetic preference, and these two effects need not be inferred from one another. In the present case, increasing $\omega$ raises the potential at fixed separation, while the point at which $V_\parallel$ crosses zero shifts toward smaller separations. Increasing the corotating temperature has the same qualitative effect on the zero of the potential. The magnitude of the rotational modification of the potential, however, becomes smaller as $T_{\rm cor}$ is increased.

It is also useful to examine the bulk location of the configuration with maximal separation. As shown in the left panel of Fig.\,\ref{maxL}, $z_{c,\parallel}^{\rm max}$ increases with $\omega$. Thus, while the maximum boundary separation decreases, the turning point of the maximally extended string moves deeper into the bulk. Notice that the behavior of $z_{c,\parallel}^{\rm max}$ follows the same qualitative trend as the position of the event horizon $z_h$, which also moves to larger values of $z$ with increasing $\omega$ at fixed $T_{\rm cor}$. The corresponding maximal separation decreases with increasing angular velocity over most of the deconfined range. Close to the endpoint of the allowed rotational range, a small upward bending is observed, particularly at higher $T_{\rm cor}$. Since this feature occurs in the immediate vicinity of the phase boundary, we do not regard it as a robust reversal of the overall rotational trend. Our analysis at other temperatures above the transition temperature confirms that the overall decrease of $L_\parallel^{\rm max}$ with $\omega$ persists beyond the temperatures displayed in the figure. At fixed $\omega$, $L_\parallel^{\rm max}$ also decreases with increasing $T_{\rm cor}$. The behavior of $z_{c,\parallel}^{\rm max}$ and $L_\parallel^{\rm max}$ therefore provides complementary information about the geometric response of the connected string. For the higher temperatures considered in our analysis, the rotational effect on both $z_{c,\parallel}^{\rm max}$ and $L_\parallel^{\rm max}$ becomes weaker away from the endpoint of the allowed rotational range.

We now repeat the longitudinal Wilson-loop analysis in the static frame of the same rotating geometry. In this description, the quark--antiquark pair is taken to be static with respect to the static observer. This comparison allows us to examine how the longitudinal string configurations and their associated observables change when the rotating plasma is characterized in the two different frames.

\begin{figure}[h]
\begin{center}
\includegraphics[width=7.1cm]{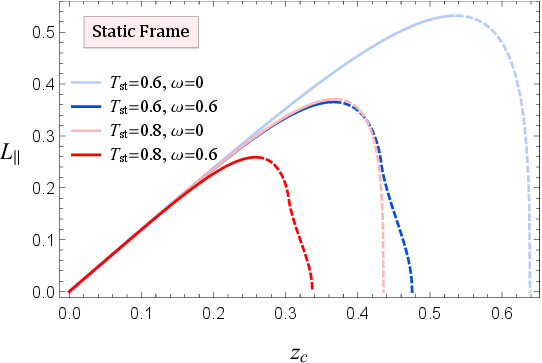}\hspace{.3cm}
\includegraphics[width=7.1cm]{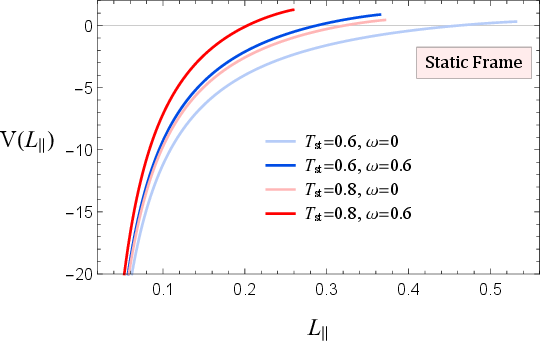}
\end{center}
\caption{\footnotesize
The longitudinal quark--antiquark separation $L_\parallel$ as a function of the turning point $z_c$ (left) and the corresponding
heavy-quark potential $V(L_\parallel)$ (right) in the static frame for $T_{\rm st}=0.6$ and $0.8~{\rm GeV}$ and for $\omega=0$ and
$0.6~{\rm GeV}$.}
\label{LV-L-st}
\end{figure}

The left panel of Fig.\,\ref{LV-L-st} shows the resulting longitudinal separation for $T_{\rm st}=0.6$ and $0.8~{\rm GeV}$ and for
$\omega=0$ and $0.6~{\rm GeV}$. The qualitative behavior is the same as in the corotating frame: increasing $\omega$ decreases
$L_\parallel$ at fixed $z_c$ and reduces the maximal separation. The rotational modification, however, is more pronounced in the static frame for the same numerical values of the temperature and angular velocity. The second, thermodynamically disfavored branch of the static-frame solution, corresponding to larger values of $z_c$, also exhibits an important geometrical feature. As the turning point approaches the region where $g_{tt}^{\rm st}=0$, this branch develops a characteristic curved behavior before eventually reaching the event horizon. This differs from the transverse configuration discussed below, for which the vanishing of $g_{tt}^{\rm st}$ produces a genuine worldsheet horizon and the corresponding transverse separation terminates at $z_*<z_h$. The distinction arises from the nonvanishing $g_{t\psi}^{\rm st}$ term in the longitudinal induced metric.

The static-frame heavy-quark potential is shown in the right panel of Fig.\,\ref{LV-L-st}. As in the corotating frame, increasing $\omega$ raises the potential at fixed separation. Increasing $T_{\rm st}$ has the same qualitative effect. The potential crosses zero at a smaller separation as either $\omega$ or $T_{\rm st}$ is increased. Comparing the two frames shows that the rotational increase of the potential at small separations is more pronounced in the corotating frame, whereas at larger separations the increase in the static-frame potential becomes stronger. Consequently, the zero of the static-frame potential occurs at a smaller separation. Thus, although the potential has the same qualitative dependence on $\omega$ in the two descriptions, its quantitative behavior is frame dependent.

\begin{figure}[h]
\begin{center}
\includegraphics[width=7.1cm]{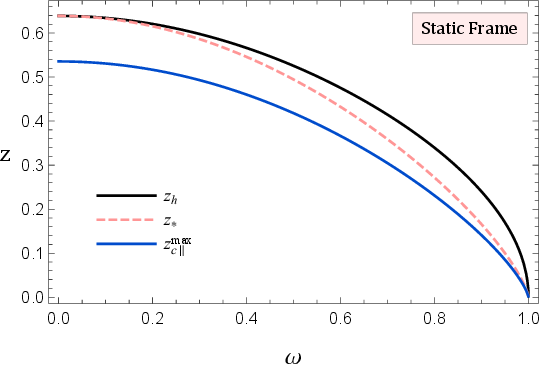}\hspace{.3cm}
\includegraphics[width=7.1cm]{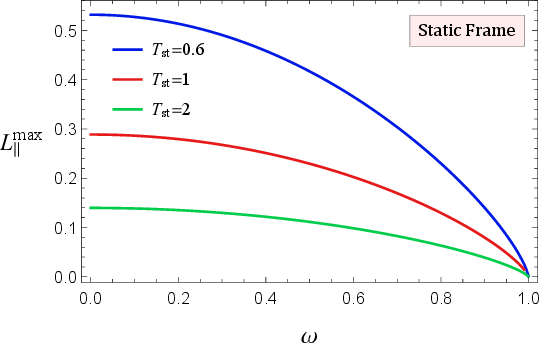}
\end{center}
\caption{\footnotesize
Left: The turning point $z_{c,\parallel}^{\rm max}$ of the maximally extended longitudinal string, the event-horizon position $z_h$, and the zero of the static-frame metric component $g_{tt}^{\rm st}$, denoted by $z_*$, as functions of the angular velocity $\omega$ for a fixed static-frame temperature. Right: The maximum longitudinal separation $L_\parallel^{\rm max}$ as a function of the angular velocity $\omega$ in the static frame for three different values of $T_{\rm st}$.}
\label{maxLst}
\end{figure}

The dependence of the maximal longitudinal separation on rotation in the static frame is displayed in the right panel of Fig.\,\ref{maxLst}. At each fixed $T_{\rm st}$, $L_\parallel^{\rm max}$ decreases monotonically with $\omega$ and tends to zero as $\omega$ approaches the causal limit. At higher $T_{\rm st}$, the maximal separation is smaller at a given $\omega$, while its decrease with angular velocity is slower. This behavior is qualitatively the same as that found in the corotating frame, although the static-frame curves remain accessible over the whole permitted causal range of the angular velocity because, unlike in the corotating frame, the critical temperature decreases with increasing $\omega$ in the static frame. Thus, once the nonrotating system is in the deconfined phase at a fixed $T_{\rm st}$, increasing the angular velocity does not drive it into the confined phase within the permitted range.

The turning point of the maximally extended string $z_{c,\parallel}^{\rm max}$ in the static frame at fixed $T_{\rm st}=0.6~{\rm GeV}$ is shown by the solid blue curve in the left panel of Fig.\,\ref{maxLst}. As can be seen, it moves toward the boundary at $z=0$ as $\omega\to1$. Figure \ref{maxLst} also displays the event horizon $z_h$ and the location $z_*$ at which the static-frame metric component $g_{tt}^{\rm st}$ vanishes, at the same temperature. At $\omega=0$, the geometry is nonrotating and $z_*$ coincides with the event horizon, as expected. Once rotation is introduced, $z_*$ separates from $z_h$ and moves toward smaller values of $z$. In contrast, in the corotating frame $g_{tt}$ vanishes at the event horizon, so that no analogous surface appears outside the horizon. The longitudinal maximal-separation turning point remains below the event horizon and follows the same decreasing trend as $z_h$ with increasing angular velocity. All three characteristic radial scales eventually approach zero as $\omega\to1$.

The comparison between the two frames therefore reveals an important distinction. Both descriptions predict a decrease of the longitudinal separation and of $L_\parallel^{\rm max}$ with increasing angular velocity, and both predict an increase of the heavy-quark potential at fixed separation. Hence, the longitudinal Wilson-loop observables do not exhibit a reversal of their qualitative $\omega$ dependence when the quark pair is taken to be static in the static frame rather than in the corotating frame. Nevertheless, their quantitative behavior is different: the magnitude of the rotational correction to the separation is larger in the static frame, while the potential shows a separation-dependent difference between the two descriptions. The location of the zero of the potential shifts toward smaller separations with increasing $\omega$ and with increasing temperature, although the shift associated with rotation is relatively modest in the corotating frame compared with the static-frame result.

%==========================================================================
\subsection{Transverse Wilson loop and heavy-quark potential}

We next consider the transverse configuration, for which the string is extended along the $\theta$ direction. This direction is orthogonal to the longitudinal $\psi$ direction at arbitrary $\theta$, and the corresponding physical separation is denoted by $L_\perp$. We first discuss the results in the corotating frame and then compare them with the corresponding static-frame configurations.

\begin{figure}[h]
\begin{center}
\includegraphics[width=7.1cm]{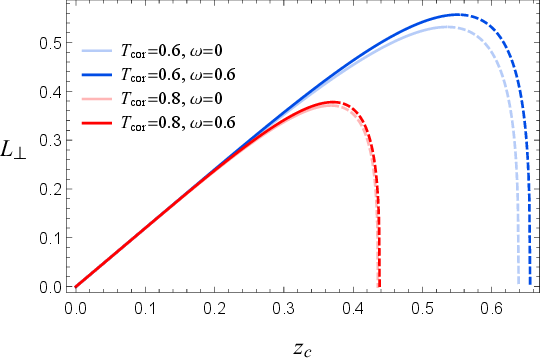}\hspace{.3cm}
\includegraphics[width=7.1cm]{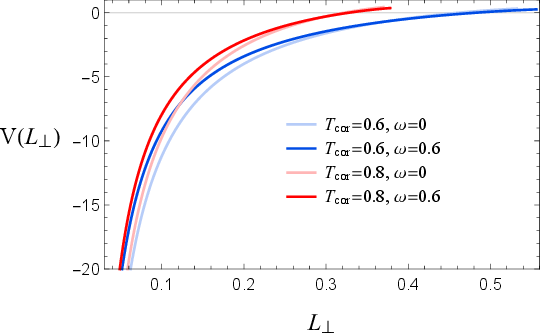}
\end{center}
\caption{\footnotesize
The transverse quark--antiquark separation $L_\perp$ as a function of the turning point $z_c$ (left) and the corresponding heavy-quark potential $V(L_\perp)$ (right) for two values of $T_{\rm cor}$ and for $\omega=0$ and $0.6~{\rm GeV}$ in the corotating frame.}
\label{LV-T}
\end{figure}

The left panel of Fig.\,\ref{LV-T} displays $L_\perp(z_c)$ for the same values of $T_{\rm cor}$ and $\omega$,  as in the longitudinal direction case. The two connected branches and their thermodynamic interpretation are analogous to those found in the
longitudinal case. In contrast to the longitudinal configuration, however, increasing $\omega$ shifts the transverse separation to larger values at fixed $z_c$ and increases the maximal separation. Thus, in the corotating frame, the rotational response of the transverse separation is opposite to that found for the longitudinal configuration.

The corresponding potentials are shown in the right panel of Fig.\,\ref{LV-T}. As in the longitudinal case, rotation raises the potential at fixed separation. The connected potential is negative at sufficiently small separation and approaches zero as the separation increases. Its zero therefore marks the point at which the connected and disconnected configurations have equal energy, and hence the connected configuration ceases to be energetically preferred over the disconnected one. The increase of the potential with rotation is accompanied, however, by an increase of the maximal separation. Thus, the rotational modification of the energy of a connected configuration and the change in the range over which such configurations exist are distinct effects.

\begin{figure}[h]
\begin{center}
\includegraphics[width=7.1cm]{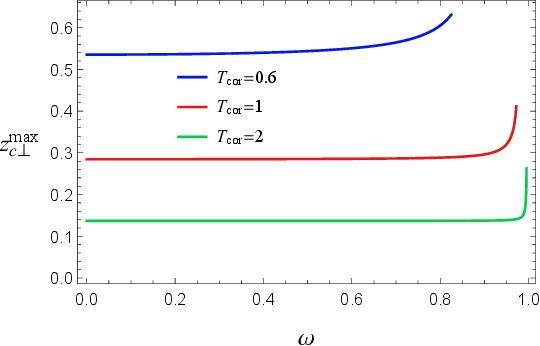}\hspace{.3cm}
\includegraphics[width=7.1cm]{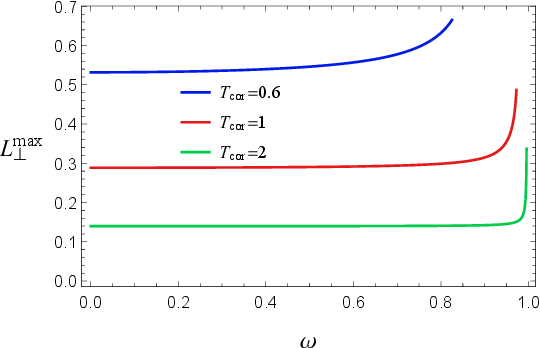}
\end{center}
\caption{\footnotesize
The turning point $z_{c,\perp}^{\rm max}$ corresponding to the maximum transverse separation (left) and the maximum transverse separation $L_\perp^{\rm max}$ (right) as functions of the angular velocity $\omega$ for three values of the corotating  temperature.}
\label{maxT}
\end{figure}

The maximal-separation configurations provide a further distinction between the two orientations in the corotating frame. As displayed in Fig.\,\ref{maxT}, $z_{c,\perp}^{\rm max}$ increases with $\omega$, following the same qualitative trend as the event-horizon position $z_h$, while $L_\perp^{\rm max}$ also increases with $\omega$. Hence, the deeper bulk location of the maximally extended string is associated with opposite changes in the boundary separation in the two orientations. Comparing the results at different temperatures shows that the rotational modification becomes weaker as $T_{\rm cor}$ is increased, which is evident in the rate of changes in $L_\perp^{\rm max}(\omega)$ at various temperatures. The effect of rotation is therefore most visible at lower temperatures, while the qualitative distinction between the longitudinal and transverse responses remains unchanged at higher temperatures.

We now repeat the transverse Wilson-loop analysis in the static frame of the same rotating geometry. As in the longitudinal case, the quark--antiquark pair is static with respect to the static observer. In the static frame, the transverse Wilson-loop observables are found to be very close to their longitudinal counterparts, although small orientation-dependent differences remain at finite angular velocity.

\begin{figure}[h]
\begin{center}
\includegraphics[width=7.1cm]{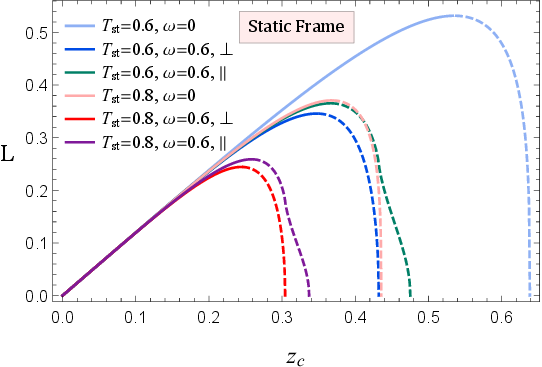}\hspace{.3cm}
\includegraphics[width=7.1cm]{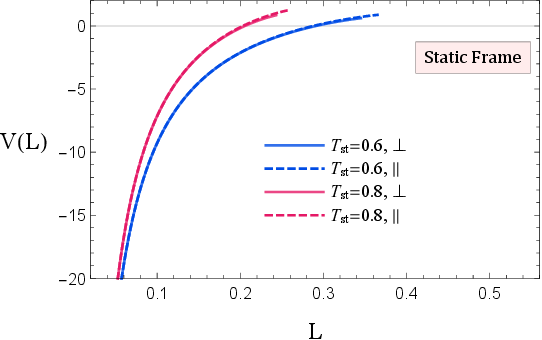}
\end{center}
\caption{\footnotesize
Static-frame comparison of the longitudinal and transverse Wilson-loop observables. Left: the quark--antiquark separation $L(z_c)$ for $T_{\rm st}=0.6$ and $0.8~{\rm GeV}$ and for $\omega=0$ and $0.6~{\rm GeV}$. Right: Heavy-quark potential in the static frame for the longitudinal and transverse configurations at $T_{\rm st}=0.6$ and $0.8~{\rm GeV}$ and $\omega=0.6~{\rm GeV}$. Solid and dashed curves correspond to the transverse and longitudinal configurations, respectively.}
\label{LV-LT-st}
\end{figure}

The left panel of Fig.\,\ref{LV-LT-st} compares the longitudinal and transverse separations as functions of $z_c$ at 
$T_{\rm st}=0.6$ and $0.8~{\rm GeV}$. In the absence of rotation, the two orientations coincide, as required by the isotropy of the
nonrotating geometry. At finite angular velocity, both separations decrease with $\omega$ at fixed $z_c$, and the magnitude of this
decrease is very close in the two orientations, although it is slightly larger for the transverse configuration. The second branch of
the transverse solution exhibits, however, a distinctive behavior: unlike the longitudinal branch, it terminates at $z_*$, where
$g_{tt}^{\rm st}=0$. This is the worldsheet horizon of the transverse configuration in the static frame. The longitudinal branch does not terminate there because its induced metric contains the nonvanishing $g_{t\psi}^{\rm st}$ term and instead continues to the event horizon.

The corresponding static-frame potentials for $T_{\rm st}=0.6$ and $0.8~{\rm GeV}$ at $\omega=0.6~{\rm GeV}$ are compared in the right panel of Fig.\,\ref{LV-LT-st}. At fixed separation, rotation increases the potential in both orientations, as already found for the longitudinal configuration. The magnitude of this rotational increase is nearly identical for the two orientations, but is slightly smaller in the transverse case. Consequently, at a given $T_{\rm st}$ and nonzero $\omega$, $V(L_\perp)$ lies slightly below $V(L_\parallel)$. 

\begin{figure}[h]
\begin{center}
\includegraphics[width=7.1cm]{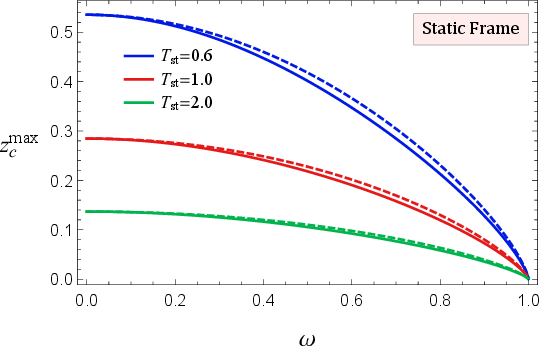}\hspace{.3cm}
\includegraphics[width=7.1cm]{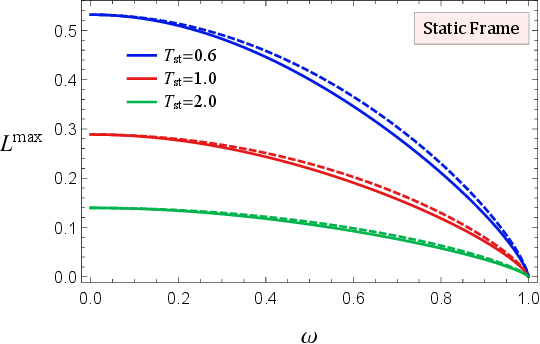}
\end{center}
\caption{\footnotesize
Static-frame comparison of the maximal-separation configurations for the longitudinal and transverse orientations. The left panel shows the turning point $z_c^{\rm max}$ and the right panel the maximum separation $L^{\rm max}$ as functions of $\omega$ for three values of $T_{\rm st}$. Solid and dashed curves correspond to the transverse and longitudinal configurations, respectively.}
\label{maxLT-st}
\end{figure}

The close correspondence between the two orientations in the static-frame description is also reflected in the turning points of
the maximally extended strings. As shown in the left panel of Fig.\,\ref{maxLT-st}, $z_{c,\perp}^{\rm max}$ follows the same
decreasing behavior with $\omega$ as $z_{c,\parallel}^{\rm max}$ and remains slightly smaller than $z_{c,\parallel}^{\rm max}$ at finite angular velocity. The right panel shows the corresponding behavior of $L^{\rm max}$ and makes the small but systematic difference between the two orientations more apparent. Thus, although the transverse separation is more strongly
suppressed by rotation than the longitudinal separation, the orientation dependence of the static-frame maximal separation remains quantitatively small. In both cases, the longitudinal and transverse quantities coincide at $\omega=0$, where rotational anisotropy is absent, and approach the same limiting value as $\omega\to1$.  The difference between the two orientations becomes again progressively smaller as $T_{\rm st}$ is increased. 

The comparison of the two orientations in the static frame is noteworthy when contrasted with the corotating results. In the corotating frame, the effective string tensions, the separation, and the maximal separation exhibit clear qualitative differences between the longitudinal and transverse configurations, whereas in the static frame these Wilson-loop observables have the same qualitative dependence on $\omega$ and differ only in magnitude. The transverse separation is suppressed slightly more strongly than the longitudinal one, while the transverse potential is increased slightly less strongly by rotation. Thus, the anisotropic response of the Wilson loop is substantially stronger in the corotating frame, both qualitatively and quantitatively. The static frame exhibits only a small quantitative anisotropy in the Wilson-loop observables, in contrast to the pronounced orientation dependence found when the quark--antiquark pair is static in the corotating frame. Together with the effective-tension results of Sec.~4.1, this comparison shows that the choice of frame does not simply reverse the rotational effects; rather, it changes both their magnitude and, for some observables, their qualitative orientation dependence.

%*********************************************************************************************************************
\section{Discussion and concluding remarks}

In this work, we have investigated the heavy-quark sector of a rotating strongly coupled plasma using an equal-rotation Myers--Perry black hole in asymptotically global AdS. Our main aim was to examine how the frame in which the quark--antiquark pair is taken to be static affects the rotational response of Wilson-loop observables. In particular, we considered two descriptions of the same rotating geometry: a corotating frame, in which the pair is static with respect to the rotating plasma, and a static frame, in which the pair is static with respect to the nonrotating observer. These two constructions therefore correspond to different physical configurations of the quark pair, rather than two observations of the same static configuration. In each frame, the analysis was performed at fixed temperature measured in that frame, allowing the effect of rotation to be studied at fixed thermodynamic conditions.

In the corotating frame, we find a pronounced orientation dependence of the Wilson-loop observables. The longitudinal and transverse effective string tensions respond differently to increasing angular velocity, with either decreasing or increasing behavior depending on the orientation and the radial position in the bulk. This distinction persists in the separation and maximal separation of the connected string configurations: rotation suppresses the longitudinal separation while enhancing the transverse one. The heavy-quark potential, in contrast, increases with angular velocity in both orientations. Thus, the effect of rotation cannot be characterized by a universal enhancement or suppression of the string dynamics; both the orientation of the string and its position in the bulk play an important role.

The static-frame analysis reveals a markedly different pattern. The longitudinal and transverse effective string tensions both decrease with angular velocity and vanish at finite angular velocities within the causally allowed range, although the mechanisms associated with their vanishing are different. The Wilson-loop observables in the two orientations also exhibit the same qualitative rotational dependence: both the separation and the maximal separation decrease with angular velocity, while the heavy-quark potential increases. The transverse effects are slightly stronger for the separation and maximal separation, whereas the longitudinal potential shows a somewhat stronger rotational increase. The difference between the two orientations in the static frame is therefore mainly quantitative, in contrast to the pronounced qualitative and quantitative anisotropy found in the corotating frame.

The present analysis provides a direct illustration of the importance of the observer choice in rotating holographic systems. Although the underlying rotating geometry is the same, the Wilson loop constructed for a pair static in the corotating frame exhibits a substantially stronger orientation dependence than that obtained for a pair static in the static frame. In particular, rotation changes the qualitative behavior of the separation and maximal separation between the two orientations in the corotating frame, whereas these quantities are both suppressed in the static frame. The heavy-quark potential retains the same qualitative increase with angular velocity in all cases, but its magnitude and its dependence on the orientation and separation are also frame dependent. These results show that the choice of observer used to define a static quark--antiquark configuration in a rotating plasma can substantially modify the rotational response of the heavy-quark observables and specifies the physical configuration represented by the Wilson loop.

It is also useful to relate our results to the recent lattice study of static quark--antiquark interactions in a rotating gluonic plasma \cite{lattice2026}. In that work, the rotating system is formulated in a corotating frame, as is natural for the lattice implementation of the
rotating medium, and static quark--antiquark interactions are studied for both longitudinal and transverse configurations. Although the lattice and holographic constructions differ in their underlying dynamics and in the precise definitions of the observables, the treatment of the rotating
medium and the static sources provides a close counterpart to our corotating setup. The weakening of the rotational modification of the
quark--antiquark interaction with increasing temperature found in the lattice calculation \cite{lattice2026} is also consistent with the temperature dependence observed in our corotating results, as well as with the behavior found in other holographic descriptions of rotating plasmas. A more direct comparison would require holographic calculations formulated with thermodynamic and rotational parameters closer to those of the lattice setup. This is particularly relevant for future lattice studies at real angular velocity, which would provide an opportunity for a
more quantitative comparison with holographic calculations. Moreover, extending the present analysis to imaginary angular velocity would provide a natural connection to the current lattice calculation, which is performed at imaginary angular velocity.

Several extensions of the present analysis would therefore be worthwhile. A natural next step is to investigate other nonlocal probes, such as the
Schwinger effect and entanglement observables, in the same static and corotating descriptions, and to determine whether the frame-dependent
anisotropy found here persists. It would also be valuable to extend the analysis to holographic constructions in which rotation is generated by a rotating boundary gauge field, both with and without backreaction on the gluodynamics. More generally, comparisons with other established rotating holographic backgrounds, including boost-generated and globally rotating constructions, under matched thermodynamic conditions could help separate effects associated with the physical rotation of the plasma from those arising from the observer choice and from the particular holographic realization of rotation.

%*********************************************************************************************************************
%\section*{Acknowledgement}
%\appendix
%\section*{Appendix A} \label{Calculation}
%\setcounter{equation}{0}
%\renewcommand{\theequation}{\Alph{section}.\arabic{equation}}

 \end{document}